# Observations on Unstable Quantons[1], Hyperplane Dependence and Quantum Fields

Gordon N. Fleming*

*Prof. Emeritus, Physics, Pennsylvania State Univ.*

* gnf1@psu.edu , gnf1@earthlink.net

There is persistent heterodoxy in the physics literature concerning the proper treatment of those quantons that are unstable against spontaneous decay. Following a brief litany of this heterodoxy, I develop some of the consequences of assuming that such quantons can exist, undecayed and isolated, at definite times and that their Lorentz covariant treatment can be carried out within a standard quantum theoretic state space. This assumption requires hyperplane dependence for the unstable quanton states and leads to clarification of some recent results concerning deviations from relativistic time dilation of decay lifetimes. In the course of the discussion I make some observations on the relationship of unstable quantons to quantum fields.

**1. Introduction:** The recent chorus against the compatibility of the whole quanton conceptual apparatus with quantum field theory remains strong, if not rising [Da 84], [Wa 94], [Hal 02], [Fr 08]. If stable quantons can be driven from the elysian quantum fields, can unstable quantons fare any better? I am not yet convinced by the arguments against quantons in QFT[2]. But should I become so, then, it seems to me, so much the worse for QFT! For the primary support for the quanton based conceptual apparatus does not lie within refined theory but within robust phenomenology! The vast wealth of empirical data concerning the world of quantum phenomena is overwhelmingly well represented within the quanton conceptual scheme[3]. We are not likely to abandon quantons due to the subtle mathematics of unitarily inequivalent representations or functional analysis in curved space-time so long as we continue to observe quanton 'tracks' and 'traces' in and by apparatus. Therefore I take heart from the efforts to upgrade the quanton conceptual apparatus, via distinctions between the traditional, global quanton concepts and *regionally* local quanton concepts [Co 09] and especially for use in some emerging forms of QFT that may be required in quantum gravity research [Oe 06].

But these discussions are couched in terms of *stable* quantons and quantum fields, while *unstable* quantons have never enjoyed unanimity concerning their proper treatment in themselves, let alone their relationship to quantum fields. Is this a matter for concern?

---

1. I follow Levy-Leblond [Le 88] and others in referring to the molecular, atomic, nuclear and sub-nuclear constituents of the mass-energy world as quantons rather than particles.
2. In particular, the arguments seem dependent on somewhat artificial forms of QFT that do not recognize the prevailing conviction that QFT is an *effective theory* rather than a fundamental one [Fl 02], [Ge 93].
3. This is an appropriate context in which to emphasize the contrast between the 'quanton' terminology employed here, and the traditional 'particle' terminology with its insidious, prejudicial, classical baggage. Perhaps the field theorists are just driving out particles rather than quantons. In that case, I'm in their camp!

*There are far more unstable quantons than stable ones!* There are many more excited states than ground states, many more radioactive nuclei than not, and among the sub-nuclear quantons, only the photon, electron, neutrinos and their anti-quantons are stable. We haven't seen the proton decay yet, but theory seems to demand it.

Notwithstanding this commonality of instability, establishing the place of unstable quantons within quantum theory, especially Lorentz covariant quantum theory, to everyone's satisfaction, has proved elusive. While the earliest contributions to decay theory, [Di 27], [Ga 28], [We 30], [Fe 34], had led to some semblance of orthodoxy by mid 20th century, [Hei 53], [Hö 58], [Go 64a], the 60s saw the emergence of ambiguities in securing that orthodoxy that have not been resolved to the present day.

All the ambiguities are linked to some degree and I present here a selection with sometimes overlapping references: Just what is the status of the exponential decay that tends to dominate the temporal evolution of (the longer lived) unstable quantons? Is it just a good approximation or is there something more fundamental involved? [Kh 58] , [Lé 59b], [Win 61], [Ne 61], [Go 64b], [Te 72], [Fl 73], [Fo 73, 78], [Pe 80], [Bo 81]? What are the appropriate group or semi-group representations of the Inhomogeneous Lorentz Group (IHLG) that should be employed to classify and characterize the state spaces of unstable quantons [Zw 63], [Sc 70, 72], [Si 70], [Wil 71], [Be 65], [Tr 75], [Ex 83], [Bo 97, 00]? How secure are the spin assignments to unstable quantons? Does their transient existence spread their spin spectrum as it spreads their mass spectrum [Be 62], [Fl 72, 79]? How are the contributions of unstable quantons to scattering processes to be characterized and recognized [Go 64b], [Ca 68], [Lur 68], [Stu 97], [Bo 00, 03], [Sa 02], [Ka 08]? Do the velocity eigenstates, rather than the 3-momentum eigenstates - which are not the same for unstable quantons - form a more suitable basis for the state spaces [Zw 63], [Ha 72], [Ra 73], [Bo 00], [Se 75], [Ta 08]?  How are unstable quantons to be understood in the context of quantum field theory, [Ar 57], [Ma 58, 59], [Lé 59a], [Mc 63], [Luk 67], [Ham 72], [Li 73], [Se 75], [Stu 97], [Za 05]. As a counterpoint, efforts to accommodate unstable quantons outside of QFT appear from time to time [Ho 95], [Wic 08]. Finally, in a novel development, several workers have examined the dependence of unstable quanton lifetimes on momenta and/or velocities and noted deviations from the classically expected Einstein time-dilation result [Kh 97], [Sh 04, 08], [Heg 06],  [Ste 96, 08]. One is left with the realization that the answer to just where unstable quantons belong in Lorentz covariant quantum theory is unsettled at best.

This question will not be resolved here. In this paper I examine some consequences, for unstable quantons and their relationship to quantum field theory, that follow from the continued use of the traditional, global quanton concepts in Minkowski space-time. In particular, the presumption that unstable quantons can exist, undecayed and in isolation[4], at definite instants of time and are subject to standard state vector representation in a complex linear state space carrying a unitary representation of the full inhomogeneous Lorentz group. Of course, we never encounter in nature or in the lab, a situation in which

---

4. Thus I leave quarks and gluons aside as being, in this regard, alien species not amenable to isolation from their fellows.

*we know with certainty* that an unstable quanton of a particular species exists undecayed and alone (i.e., alone with the vacuum) in some spatial volume and at some moment of time. But we also never encounter exact momentum eigenstates or exact position eigenstates, and yet the concepts of such (improper) states are extremely valuable as idealizations and in forming bases for the expression of more realistic states. Similarly, I assume here the importance of the idealization of states for unstable quantons in which exactly one such quanton is isolated and 100% undecayed at a definite instant of time. Such states will be called **instantaneous single parent** (ISP) states. I will not examine the description of multiple quanton states using these ISP states, but only the relationships among the ISP states, themselves, and some of their relationships with quantum fields. I ignore spin in this paper.

The reader may well wonder why there should be anything new to say that could follow from such an assumption as the existence of ISP states. Surely others have presumed the possibility of instantaneous, undecayed, isolated states of unstable quantons. Indeed they have (I think almost all workers have), but it appears they have taken the notion so much for granted as to ignore it and, definitely, not take it seriously. The importance of unstable quantons demands it be taken seriously[5].

The time instant at which the probability vanishes for finding decay products will be called the **no-decay time**. At any other time the decay probability is non-zero and the probability of finding the undecayed parent quanton is less than unity. We also assume that the ISP states transform into equally accessible states under unitary representations of the full Inhomogeneous Lorentz Group (IHLG). Under the transformations of the Euclidean subgroup and/or time translations the ISP states remain ISP states with only a possible change in the no-decay time. If the transformations involve boosts, however, the ISP states transform into states which *do not have a no-decay time*. Instead, their no-decay status holds on the non-instantaneous hyperplane into which the boost transformed the original, no-decay, instantaneous hyperplane. The ISP states, together with their transforms under the IHLG will be called **single parent** (SP) states. Thus, the SP states of unstable quantons are *inherently hyperplane dependent*.

In section **2** these assumptions and the attendent concepts and necessary formalism will be developed. In particular, 3-momentum eigenstates for the ISP states will be examined and the concept of a *generalized* or **SP momentum eigenstate** will be developed for all the SP states. We will see that no SP state can have two, distinct, no-decay hyperplanes.

In section **3** we take a first look at the relation of the SP states to **quantum fields** by examining the matrix element of an *arbitrary* field between the vacuum and an SP state. For a stable quanton the corresponding matrix elements always satisfy a Klein-Gordon equation, regardless of the presence or absence of interactions. We find the *generalization of the Klein-Gordon* equation that holds for the vacuum-SP matrix

---

5. An earlier effort to take these ideas seriously [Fl 72, 79] led to questioning the standard spin assignments to unstable quantons and the development of a spin state perturbation theory.

element and we relate the matrix element to the **survival amplitude** of the unstable quanton.

In section **4** we address the concept of **lifetime** for unstable quantons. We adopt a definition of lifetime that is independent of the existence or dominance of exponential decay, [Fl 73, 78], but is compatible with the traditional evaluation if exponential decay is dominant. Then we examine the dependence of lifetime on the SP-momenta for the eigenstates of such. As a consequence of the indefinite mass of an unstable quanton, no two of these SP-momentum eigenstates, with distinct momentum eigenvalues, and *for a fixed hyperplane*, can be the Lorentz transforms of one another. Nevertheless, they do display *approximate* relativistic time dilation of lifetime with 'increasing' SP-momentum and we *explain* this. The approximate character of this dilation has drawn attention recently, [Kh 97], [Ste 08][6] and we comment on it.

In section **5** we turn from 3-momentum eigenstates to **velocity** eigenstates which *do* transform into one another under boosts and have repeatedly been championed as providing a basis superior to the 3-momentum basis for unstable quantons [Zw 63], [Ham 72], [Ra 73], [Bo 97, 00, 03], [Ta 08]. But again a recent study has argued the lifetime of such velocity eigenstates to *contract* rather than dilate under boost induced increases in speed [Sh 08]. We examine this and *explain* it. In the process we show that the velocity eigenstates comprise, in fact, a severely restricted subset of the entire collection of SP-momentum eigenstates for arbitrary hyperplanes. As such we argue that the velocity eigenstates are completely incapable of constituting a basis for the SP states of an unstable quanton. In particular, each distinct velocity eigenvalue corresponds to an SP state on a differently oriented hyperplane. Consequently, non-trivial superpositions of velocity eigenstates are not SP states on any hyperplane!

In section **6** we return to fields and examine the contribution to the **two point vacuum expectation value** from the subspace of states in which the SP states for an unstable quanton lie. The result bears on the question of the sense and degree in which a species of isolated unstable quantons can be said to be the quanta of a local field! In particular, we note the prospect of a simple change from local to *hyperplane dependent* field that would appear to strengthen the quanton-field connection.

Finally, in section **7** we examine the inner product between SP-momentum eigenstates with distinct eigenvalues and associated with intersecting hyperplanes. We find that such inner products, if non-zero, receive contributions from shared 4-momentum eigenstates corresponding to one and only one 4-momentum eigenvalue! This mathematically elementary result is physically surprising and we offer some interpretive commentary.

There are two appendices.

---

6. Although occasionally commented on here, direct comparisons between the present treatment and that of [Kh 97] and [Ste 08] are hard to come by and awkward.

**2. Single parent states:** We will employ the $+---$ Minkowski metric and will work in the Heisenberg picture in which state vectors represent entire unitary histories (free of state reductions) of systems. The complication of spin will be neglected here.

As described above, we assume that any unstable quanton can, in principle, exist alone and undecayed in space at any instant of time. A state vector of such a system can be labeled by the time at which the parent quanton has unit probability of existing undecayed - *the no-decay time*. These are the ISP states. An immediate consequence of this assumption is the existence of (improper) ISP states of precise 3-momentum. This follows from the invariance of the ISP character of a state under arbitrary spatial translation and the ISP character of an arbitrary superposition of ISP states having the same no-decay time.

In other words, if $|\psi;t>$ is an ISP state with the no-decay time, $t$, then so is,

$$|\psi;t,\mathbf{a}> = \exp[-(i/\hbar)\hat{\mathbf{P}}\cdot\mathbf{a}]|\psi;t>, \qquad (2.1)$$

and so is,

$$\int d^3 a \, |\psi;t,\mathbf{a}> \exp[(i/\hbar)\mathbf{k}\cdot\mathbf{a}] = \int d^3 a \exp[(i/\hbar)(\mathbf{k}-\hat{\mathbf{P}})\cdot\mathbf{a}]|\psi;t>$$

$$= (2\pi\hbar)^3 \delta^3(\mathbf{k}-\hat{\mathbf{P}})|\psi;t> = (2\pi\hbar)^3 |\mathbf{k};t>\psi(\mathbf{k}). \qquad (2.2)$$

But this last is a multiple of a 3-momentum eigenstate with momentum eigenvalue, $\mathbf{k}$. Clearly, the space of ISP states with a given no-decay time is spanned by the corresponding 3-momentum eigenstates, i.e.,

$$|\psi;t> = \int d^3 k \, |\mathbf{k};t>\psi(\mathbf{k}). \qquad (2.3)$$

These elementary results are obvious for stable quantons, but, as we will see, they have non-trivial consequences for unstable quantons and not all approaches to unstable quantons are compatible with them.

Under the group of Euclidean transformations, $\mathbf{x}' = R\mathbf{x} + \mathbf{a}$, and time translations, $t' = t + b$, these ISP states transform among themselves

$$\hat{U}(R,\mathbf{a},b)|\psi;t> = |\psi_{R,\mathbf{a}};t+b> \qquad (2.4)$$

where

$$\psi_{R,\mathbf{a}}(\mathbf{k}) = \psi(R^{-1}\mathbf{k})\exp[-(i/\hbar)\mathbf{a}\cdot\mathbf{k}]. \qquad (2.5)$$

Under Lorentz boosts, however, ISP states are converted into states in which the parent quanton is definitely undecayed on a *non-instantaneous*, space-like hyperplane. Interpreted passively, the boost changes the inertial frame perspective to one in which the original no-decay hyperplane no longer 'appears' instantaneous (**Fig. 1a**). Interpreted actively, the boost changes the state of the system, in part, by reorienting the no-decay hyperplane (**Fig. 1b**). So our basic assumption, coupled with the standard assumption of covariance under the IHLG, requires us to expand our conceptual framework and notation to include these single parent states with non-instantaneous or 'tilted' no-decay hyperplanes, which are not ISP states. The new states and the ISP states, taken together, will be called single parent (SP) states and, in all cases, we will refer to such states as being *"on"* the no-decay hyperplane.

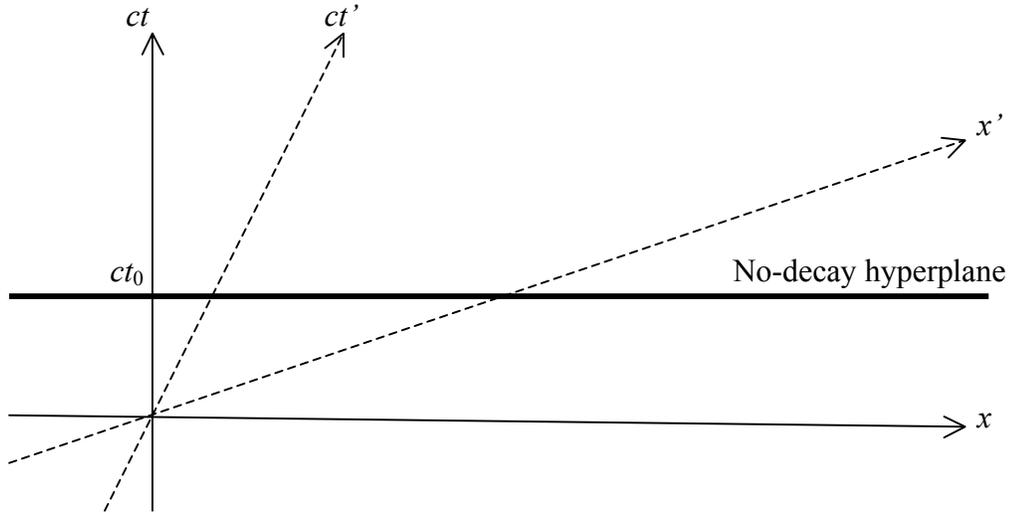

**Fig.1a:** No-decay hyperplane; instantaneous from the unprimed frame perspective, non-instantaneous from the primed frame perspective.

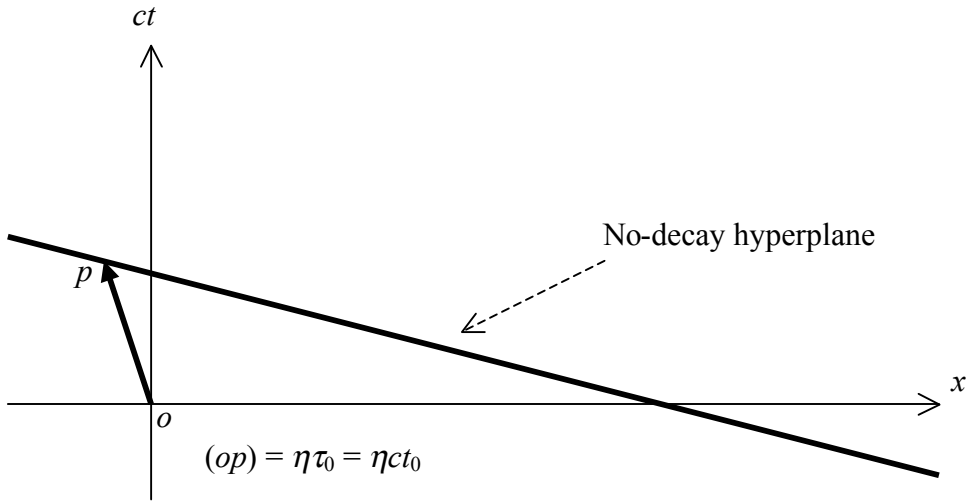

**Fig. 1b:** Non-instantaneous no-decay hyperplane with time-like normal parameterization.

The following explains the notation used here for arbitrary space-like hyperplanes. Any space-like hyperplane may be denoted by a dimensionless, future pointing, time-like unit 4-vector, $\eta$, which is orthogonal to the hyperplane, and a parameter, $\tau$, such that, from the perspective of an inertial frame employing Minkowski coordinates, $x$, any point on the hyperplane has coordinates satisfying,

$$x^\mu \eta_\mu = \tau. \tag{2.6}$$

The special case of instantaneous hyperplanes is given by $\eta^{(0)} = (1, \mathbf{0})$. Any boost, $B$, that takes $(1, \mathbf{0})$ into $\eta = (\eta^0, \boldsymbol{\eta})$ will also take the '3-momentum' $(0, \mathbf{k})$ into $p = (p^0, \mathbf{p})$ where $\eta p = 0$.

Accordingly, for a boost, $B$, we will write,

$$\hat{U}(B)|\psi;t> = \hat{U}(B)|\psi;\eta^{(0)},ct> = |\psi_B;\eta,\tau> \tag{2.7}$$

where, $ct = \tau$, and

$$\psi_B(p) = \psi(B^{-1}p) = \psi((0,\mathbf{k})) = \psi(\mathbf{k}). \tag{2.8}$$

Similarly and finally, for any inhomogeneous Lorentz transformation, $(L, a)$, we have,

$$\hat{U}(L,a)|\psi;\eta,\tau> = |\psi_{L,a};L\eta,\tau+aL\eta>, \tag{2.9}$$

where,

$$\psi_{L,a}(p) = \psi(L^{-1}p)\exp[(i/\hbar)pa]. \tag{2.10}$$

The SP-momentum eigenstates contributing to an SP state vector can be obtained, in analogy with (2.2), by the appropriate superposition of space-like translations (parallel to the no-decay hyperplane) of the original SP state vector. Thus we have,

$$\int \frac{d^4\lambda \delta(\eta\lambda)}{(2\pi\hbar)^3} \exp[(i/\hbar)\lambda(\hat{P}-p)]|\psi;\eta,\tau> = \delta^3_\eta(\hat{P}-p)|\psi;\eta,\tau>$$

$$= |p;\eta,\tau>\psi(p), \tag{2.11}$$

where,

$$<p';\eta,\tau|p;\eta,\tau> = \delta^3_\eta(p'-p), \tag{2.12}$$

and,

$$\delta^3_\eta(q) := (2\pi\hbar)^{-3}\int d^4\lambda \delta(\eta\lambda)\exp[(i/\hbar)\lambda q] = \frac{1}{\eta^0}\delta^3(\mathbf{q} - \frac{\vec{\eta}}{\eta^0}q^0), \tag{2.13}$$

vanishes if $q$ is not parallel to $\eta$. The inverse of (2.11), the expansion of the SP state vector in terms of the SP-momentum eigenstates, is given by,

$$|\psi;\eta,\tau> = \int d^4p\, \delta(\eta p)|p;\eta,\tau>\psi(p) := \int d^3_\eta p\, |p;\eta,\tau>\psi(p). \tag{2.14}$$

From the invariance of the mass operator, $\sqrt{\hat{P}^2}$, under the IHLG, it follows that the mass spectrum distribution function, $\sigma_{p^2}(\mu)$, for SP momentum eigenstates, defined by,

$$\delta_\eta^3(p'-p)\sigma_{p^2}(\mu) := \,<p';\eta,\tau\,|\,\delta(\mu-\sqrt{\hat{P}^2}\,)\,|\,p;\eta,\tau> \quad , \tag{2.15}$$

can not depend on $\eta$ or $\tau$, but can depend on the invariant, $p^2$. To admit such dependence, however, is to risk undermining the concept of an unstable quanton of any content distinct from an arbitrary superposition of asymptotic decay product states. Speaking broadly, theorists tend to deny such dependence while phenomenologists tend to admit it [Ja 64]. Dynamical arguments can be mounted for denying it[7] and we will deny such dependence here.

Consequently, *all* SP states for a given species of unstable quanton have the same distribution function over the mass spectrum, i.e., for unit norm $\psi$,

$$\sigma(\mu) := \,<\psi;\eta,\tau\,|\,\delta(\sqrt{\hat{P}^2}-\mu)\,|\,\psi;\eta,\tau>, \tag{2.16}$$

where $\sigma(\mu)$ is independent of $\psi$, $\eta$, and $\tau$. This also tells us that no SP state can have two, intersecting, no-decay hyperplanes. In other words, for $\eta \ne \eta'$, we must have,

$$|\,\psi';\eta',\tau'> \,\ne\, |\,\psi;\eta,\tau>, \tag{2.17}$$

for any $\psi$, $\psi'$, $\tau$, and $\tau'$. The reason is that since,

$$d^4q = d_\eta^3 q\, d(\eta q) = \frac{d_\eta^3 q}{2\eta q}\,2\eta q\, d(\eta q) = \frac{d_\eta^3 q}{2\eta q}\, d(q^2) = d_\eta^3 p\, d(\sqrt{q^2})\left(\frac{\sqrt{q^2}}{\eta q}\right), \tag{2.18}$$

equality in (2.17) would require the common probability distribution, $P(q)$, for contributing 4-momenta, $q$, to satisfy,

$$\eta' q\,|\,\psi'(q-\eta'(\eta' q))\,|^2\,\frac{\sigma(\sqrt{q^2})}{\sqrt{q^2}} = P(q) = \eta q\,|\,\psi(q-\eta(\eta q))\,|^2\,\frac{\sigma(\sqrt{q^2})}{\sqrt{q^2}}\,. \tag{2.19}$$

But this can not hold for *all* $q$ for *any* two $\psi$ and $\psi'$. Further comment on this point will be made in **4** just after (4.10).

---

7. For example, if the dynamical acceleration of an unstable quanton with $p = 0$, by transmission of an impulse (application of a given force for a given time), is to yield the 'same' unstable quanton with $p =/= 0$, then, since the impulse transmission will not disturb the original mass spectral function, the spectral function can not depend on $p$.

**3. The quantum field connection; a first look:** In the following I will have occasion to use the useful relation [Fl 73],

$$\hat{A}|\psi> = |\psi><A>_\psi + |\psi_A>\Delta_\psi A, \tag{3.1}$$

where $\hat{A}$ is *any* operator defined on $|\psi>$, which has the same norm as $|\psi_A>$, which, in turn, is orthogonal to $|\psi>$, and where $<.>_\psi$ denotes expectation value and

$$\Delta_\psi A = \left|\sqrt{<(\hat{A}^\dagger - <A>_\psi^*)(\hat{A} - <A>_\psi)>_\psi}\right|. \tag{3.2}$$

This result generalizes naturally to continuous spectrum eigenvectors of any operator, $B$, that commutes with $A$.

We consider a quantum field, $\hat{\phi}(x)$, and examine its matrix element between the vacuum and a single quanton state vector, $<\Omega|\hat{\phi}(x)|\psi>$. If the quanton is stable, with a definite rest mass, $m$, then it is trivial and well known that regardless of the presence or nature of interactions, (and regardless of the tensorial-spinorial rank of the field) this matrix element satisfies the Klein-Gordon equation,

$$[\partial^\mu\partial_\mu + (mc/\hbar)^2]<\Omega|\hat{\phi}(x)|\psi> = 0. \tag{3.3}$$

A corresponding, but different, result holds, if the single quanton is unstable, for a vacuum – SP state matrix element of the field, $<\Omega|\hat{\phi}(x)|\psi;\eta,\tau>$. In such a case we use (3.1) in the form,

$$\hat{P}^2|\psi;\eta,\tau> = |\psi;\eta,\tau>\kappa^2\hbar^2 + |\psi_{P^2};\eta,\tau>\Delta(P^2), \tag{3.4}$$

where $\kappa^2\hbar^2 := <P^2>$, to obtain (like $\sigma$, the expectation value is independent of $\psi$),

$$[\partial^\mu\partial_\mu + \kappa^2]<\Omega|\hat{\phi}(x)|\psi;\eta,\tau> = -<\Omega|\hat{\phi}(x)|\psi_{P^2};\eta,\tau>\Delta(P^2). \tag{3.5}$$

If the matrix element on the left is non-zero then the matrix element on the right must also be non-zero if we are to avoid stable quanton evolution in $x$!

Similarly, for the SP-momentum eigenstate on the ($\eta$, $\tau$) hyperplane, $|p;\eta,\tau>$, we find (keep in mind that $p$ is orthogonal to $\eta$ and, therefore, is space-like)

$$[\hbar^2(\eta\partial)^2 - p^2 + \kappa^2\hbar^2]<\Omega|\hat{\phi}(x)|p;\eta,\tau> = -<\Omega|\hat{\phi}(x)|p_{P^2};\eta,\tau>\Delta(P^2), \tag{3.6}$$

where,

$$<\Omega|\hat{\phi}(x)|p;\eta,\tau> = <\Omega|\hat{\phi}(0)|p;\eta,\tau-\eta x>e^{-(i/\hbar)px}, \quad (3.7a)$$

and

$$<\Omega|\hat{\phi}(x)|p_{P^2};\eta,\tau> = <\Omega|\hat{\phi}(0)|p_{P^2};\eta,\tau-\eta x>e^{-(i/\hbar)px}. \quad (3.7b)$$

Now that all the x dependence is either explicit or combined with the $\tau$ dependence in the SP-momentum eigenstate, we may inquire whether it appears in the form of a decay or survival amplitude for the parent quanton? Defining the survival amplitude, $I_p(\tau)$, by[8],

$$<p';\eta,\tau'|p;\eta,\tau> = <p';\eta,\tau'|\exp[(i/\hbar)\eta\hat{P}(\tau-\tau')]|p;\eta,\tau'>$$

$$:= \delta^3_\eta(p-p')I_p(\tau'-\tau), \quad (3.8)$$

and applying (3.1) again, this time with, $\hat{A} = \exp[(i/\hbar)\eta\hat{P}\tau] := \hat{U}(\eta\tau)$, to obtain

$$\hat{U}(\eta\tau)|p;\eta,0> = |p;\eta,0>I_p(-\tau) + |p_{U(-\eta\tau)};\eta,0>\sqrt{1-|I_p(-\tau)|^2}, \quad (3.9)$$

we find,

$$<\Omega|\hat{\phi}(0)|p;\eta,\tau> = <\Omega|\hat{\phi}(0)|p;\eta,0>I_p(-\tau)$$

$$+ <\Omega|\hat{\phi}(0)|p_{U(\eta\tau)};\eta,0>\sqrt{1-|I_p(-\tau)|^2}, \quad (3.10)$$

where the square root is the rms deviation, (3.2), of the unitary operator in the eigenstate, $|p;\eta,0>$.

Note that the state, $|p_{U(\eta\tau)};\eta,0>$, is not only orthogonal to $|p;\eta,0>$, but to $|p';\eta,0>$ for any $p'$ orthogonal to $\eta$, because it is, itself, an eigenstate of, $\hat{P}^\mu - \eta^\mu(\eta\hat{P})$, with eigenvalue, $p$. Thus it is orthogonal to *all* SP states on the $(\eta,0)$ hyperplane and, on that hyperplane, consists solely of decay products.

So our field matrix element contains some information concerning the decay of the unstable quanton.

**4. Survival amplitudes and lifetime dilation:** In analogy with (3.8), a survival amplitude for a unit norm SP state, $|\psi;\eta,\tau>$, can be defined as,

$$I_{\psi,\eta}(\tau) = <\psi;\eta,\tau|\psi;\eta,0> \quad (4.1)$$

---

8. As is made explicit in (4.11) below, the survival amplitude does not depend on $\eta$ except through the constraint that $p$ must be orthogonal to $\eta$.

and from (2.11) and (3.8) we [would then] have

$$I_{\psi,\eta}(\tau) = \int d^4 p\, \delta(\eta p) |\psi(p)|^2\, I_p(\tau). \tag{4.2}$$

It is important to note, however, that the unitary evolution equation for a finite norm SP state, $|\psi;\eta,\tau>$, which can be written in the form,

$$|\psi;\eta,0> = |\psi;\eta,\tau> I_{\psi,\eta}(\tau) + |\psi_{U(-\eta\tau)};\eta,\tau> \sqrt{1 - |I_{\psi,\eta}(\tau)|^2}\ , \tag{4.3}$$

has a somewhat different interpretation than the corresponding equation for $|p;\eta,\tau>$,

$$|p;\eta,0> = |p;\eta,\tau> I_p(\tau) + |p_{U(-\eta\tau)};\eta,\tau> \sqrt{1 - |I_p(\tau)|^2}\ . \tag{4.4}$$

In the latter equation (essentially (3.9) except for the exchange $\tau \leftrightarrow 0$ in the SP states) the second term is orthogonal, due to being an eigenstate of $\hat{P}^\mu - \eta^\mu(\eta\hat{P})$, to the *entire SP state space* on the $(\eta, \tau)$ no-decay hyperplane. Therefore, on the $(\eta, \tau)$ hyperplane, $|p_{U(-\eta\tau)};\eta,\tau>$ consists *only* of decay products. In (4.3), however, $|\psi_{U(-\eta\tau)};\eta,\tau>$ is only guaranteed to be orthogonal to $|\psi;\eta,\tau>$, and this does not allow the interpretation of $|\psi_{U(-\eta\tau)};\eta,\tau>$ as consisting only of decay products on the no-decay hyperplane. By comparing (4.3) to (4.4), using (2.14), we find explicitly,

$$|\psi_{U(-\eta\tau)};\eta,\tau> \sqrt{1 - |I_{\psi,\eta}(\tau)|^2}$$

$$= \int d^4 p\, \delta(\eta p) \psi(p) \Big[ |p;\eta,\tau> (I_p(\tau) - I_{\psi,\eta}(\tau)) + |p_{U(-\eta\tau)};\eta,\tau> \sqrt{1 - |I_p(\tau)|^2} \Big]. \tag{4.5}$$

Only the last term under the integral consists of decay products on the $(\eta, \tau)$ hyperplane. Consequently, (4.1, 2) define an amplitude which corresponds to survival of the particular SP state, $\psi$, which amplitude can decrease in modulus not only via *decay* of the parent but also via evolution into SP states orthogonal to $\psi$. For unit norm SP states the survival *probability* against decay is given by the expectation value,

$$<\psi;\eta,0|\hat{\Pi}(\eta,\tau)|\psi;\eta,0> := \int d^4 p\, \delta(\eta p) |<p;\eta,\tau|\psi;\eta,0>|^2$$

$$= \int d^4 p\, \delta(\eta p) |I_p(\tau)\psi(p)|^2\,, \tag{4.6}$$

for which there is no amplitude and which is to be contrasted with the squared modulus of (4.2).

To examine the $\tau$ dependence of $I_p(\tau)$ or $I_{\psi,\eta}(\tau)$ further we use the expansion of the SP momentum eigenstate in terms of 4-momentum eigenstates [Ste 08]. We write the expansion in the form,

$$|p;\eta,\tau> = \int d^4q \, |q> \delta_\eta^3(q-p) r(q^2) \sqrt{2\eta q} \exp[(i/\hbar)\eta q \tau], \quad (4.7)$$

where,

$$<q'|q> = \delta^4(q'-q) \quad (4.8)$$

and the other factors yield both the mass distribution function,

$$\sigma(\mu) = 2\mu |r(\mu^2)|^2 \,, \quad (4.9)$$

and the normalization of (2.12). More detail than we need now concerning the structure of the eigenstates, $|q>$, will be provided below in section **6**.

If (4.7) is substituted into the right hand side of (2.11) and the $p$ integration performed, we then find that the probability amplitude for the 4-momentum, $q$, in the SP state, $|\psi;\eta,\tau>$, is given by,

$$\xi_\psi(q;\eta,\tau) = \psi(q-\eta(\eta q)) r(q^2) \sqrt{2\eta q} \exp[(i/\hbar)\eta q \tau]. \quad (4.10)$$

As mentioned in **2**, the important observation then follows that for $\eta' \neq \eta$, there is no choice of $\psi$ and $\psi'$ that will yield, $|\xi_{\psi'}(q;\eta',\tau')| = |\xi_\psi(q;\eta,\tau)|$, for all $q$. Hence no SP state can be an SP state on both of two intersecting hyperplanes[9]. Stefanovich makes note of essentially this fact with his observation of "decays caused by boosts" [Ste 08, p. 21][10].

Upon substituting (4.7) into (3.8), setting $\tau' = 0$ and employing (4.8), a little algebra yields,[11]

$$I_p(\tau) = \int d\mu \, \sigma(\mu) \exp[-(i/\hbar)\tau \sqrt{\mu^2 - p^2}] \quad (4.11)$$

and from this we may hope to extract the lifetime and its dependence on $p$.

However, two considerations should be kept in mind: (1) Given the contentious nature of exponential decay, it is desirable to have a definition of lifetime that does not depend on its assumption; (2) For a *fixed hyperplane orientation*, i.e., fixed, $\eta$, the SP-momentum eigenstates are not the Lorentz transforms of one another (a boost, $B$, that takes $p$ into $Bp$ also takes $\eta$ into $B\eta$). Instead, the ISP 3-momentum eigenstates are related by the unitary transformations generated by the Newton-Wigner (NW) position operator [Ste 08] while the SP-momentum eigenstates for an arbitrary fixed hyperplane are related by the unitary transformations generated by the hyperplane dependent generalization of the NW position [Fl 99]. Consequently, the dependence of the lifetime on $p$ is not, per se, the same as the dependence of lifetime on reference frame. In fact, since, just as for the survival amplitude itself, the lifetime will depend on $p$ only through the invariant, $p^2$, *the lifetime, while dependent on p, is invariant under a boost.*

---

9. The case of distinct *parallel* hyperplanes is ruled out by the quanton being unstable, not cyclic.
10. Due to the neglect of non-instantaneous hyperplanes, Stefanovich asserts that "the muon - - can be seen as a single particle by the [rest] observer and as - - three decay products by the moving observer."
11. This expression also follows from, $<p';\eta,\tau|p;\eta,0> = <p';\eta,0| \int d\mu \, \delta(\mu - \sqrt{\hat{P}^2}) \exp[-(i/\hbar)\tau \sqrt{\mu^2 - p^2}] |p;\eta,0>$, without the need for the expansion, (4.7).

Nevertheless, the definition of lifetime we adopt will display a close simulacrum of time dilation in its dependence on $p$.

For that definition of lifetime we will take an expression that, interpreted phenomenologically for an ISP state, is the average over decays of the time of decay from the time of initial preparation of the ISP state [Fl 73, 78]. Expressed in terms of the survival probability, $P_p(\tau) = |I_p(\tau)|^2$, this is given by,

$$c\mathrm{T}_p := -\int_0^\infty d\tau\, \tau \dot{P}_p(\tau) = \int_0^\infty d\tau\, P_p(\tau) \tag{4.12}$$

Of course, this quantum theoretic $P_p(\tau)$ can occasionally and briefly have a positive derivative corresponding to regeneration which complicates the interpretation of (4.12). On the other hand, to the degree the probability is a decaying exponential, to that degree (4.12) yields the standard lifetime.

Substituting (4.8) into the rightmost side and performing the $\tau$ integration first we obtain (see **Appendix 1**),

$$c\mathrm{T}_p = \pi\hbar \int d\mu\, \sigma^2(\mu) \frac{\sqrt{\mu^2 - p^2}}{\mu} \quad . \tag{4.13}$$

The last ratio, under the integral sign, is the time dilation factor for the SP-momentum, $p$, and a mass of $\mu$. The unstable quanton having a continuous mass spectrum, the time dilation factor is averaged over the spectrum and with the *squared* spectral density as the weighting function.[12]

For mass distributions sharply peaked about a mass value, $\mu_0$, the expansion,

---

12. Had we allowed a $p$ dependence in $\sigma(\mu)$ *all* resemblance to Einstein dilation of lifetime might have been lost.

$$\frac{\sqrt{\mu^2 - p^2}}{\mu} \approx \frac{\sqrt{\mu_0^2 - p^2}}{\mu_0} + \frac{(p/\mu_0)^2}{\sqrt{\mu_0^2 - p^2}}(\mu - \mu_0) - \frac{p^2[(3/2)\mu_0^2 - 2p^2]}{\mu_0[\mu_0^2 - p^2]^{3/2}}\frac{(\mu - \mu_0)^2}{\mu_0^2}, \quad (4.14)$$

indicates the momentum dependence of the dominant correction to the zeroth order time dilation. The definition of $\mu_0$ in (4.14) can be chosen to eliminate the linear term in $(\mu - \mu_0)$ when (4.14) is substituted into (4.13). Indeed, if we allow ourselves the rough approximation of a Breit-Wigner resonance form,

$$\sigma(\mu) = \frac{\Gamma/2\pi}{(\mu - \mu_0)^2 + (\Gamma/2)^2}, \quad (4.15)$$

integrated from $-\infty$ to $+\infty$, then, to lowest order in $\Gamma/\mu_0$, the IP-momentum dependence of the lifetime is given by,

$$T_p \approx T_0 \left( \frac{\sqrt{\mu_0^2 - p^2}}{\mu_0} - \frac{p^2((3/2)\mu_0^2 - 2p^2)}{4\mu_0[\mu_0^2 - p^2]^{3/2}} \left(\frac{\Gamma}{\mu_0}\right)^2 \right), \quad (4.16)$$

where, $T_0 = \hbar/\Gamma$. Note also that $\mu_0$ is defined differently than $\kappa$ in (3.4 – 6).

**5. Velocity eigenstates and time contraction:** The failure of the 3-momentum eigenstates, in the space of ISP states, to transform into one another under Lorentz boosts has, repeatedly over the years, provoked interest in velocity eigenstates, which do so transform [Zw 63], [Ham 72], [Ra 73], [Bo 97, 00, 03], [Ta 08]. However, Shirokov [Sh 08] has recently argued that such velocity eigenstates suffer lifetime *contraction* rather than dilation compared to the lifetime for zero velocity. Hegerfeldt [Heg 06] has commented on the seriousness of this result. Accordingly, we shall examine the velocity eigenstates in the light of our SP states on arbitrary hyperplanes.

By a velocity eigenstate we mean a $|\mathbf{u}\rangle$ that satisfies the equation,

$$\frac{\hat{\mathbf{P}}}{\hat{P}^0}|\mathbf{u}\rangle = |\mathbf{u}\rangle \mathbf{u}, \quad (5.1)$$

or,

$$\hat{\mathbf{P}}|\mathbf{u}\rangle = \hat{P}^0 |\mathbf{u}\rangle \mathbf{u}. \quad (5.2)$$

This definition in terms of a 3-velocity eigenvalue, $\mathbf{u}$, is easily converted into a 4-velocity form. First note that,

$$\hat{P}^2|\mathbf{u}\rangle = ((\hat{P}^0)^2 - \hat{\mathbf{P}}^2)|\mathbf{u}\rangle = (1 - \mathbf{u}^2)(\hat{P}^0)^2|\mathbf{u}\rangle. \quad (5.3)$$

Taking positive square roots of the non-negative operators on the left and the right, we have,

$$\sqrt{\hat{P}^2}\,|\mathbf{u}> = \sqrt{1-\mathbf{u}^2}\,\hat{P}^0\,|\mathbf{u}> . \tag{5.4}$$

From this and (5.2) we have,

$$\hat{P}^\mu\,|\mathbf{u}> = (\hat{P}^0, \hat{\mathbf{P}})\,|\mathbf{u}> = \frac{(1,\mathbf{u})}{\sqrt{1-\mathbf{u}^2}}\sqrt{\hat{P}^2}\,|\mathbf{u}> = v^\mu \sqrt{\hat{P}^2}\,|\mathbf{u}> , \tag{5.5}$$

with the 4-velocity eigenvalue, $v^\mu$.

For the case $\mathbf{u} = \mathbf{0}$, the velocity eigenstate, $|\mathbf{0}>$, is also a 3-momentum eigenstate with $\mathbf{p} = \mathbf{0}$. Consequently, it can be found among the SP states for unstable quantons only on the instantaneous hyperplanes. On the other hand, the Lorentz boosts of the $\mathbf{u} = \mathbf{0}$ eigenstates are velocity eigenstates with non-zero eigenvalues. To see this we note that for a boost, $B_\mathbf{u}$, of 3-velocity, $\mathbf{u}$, we have,

$$\frac{\hat{\mathbf{P}}}{\hat{P}^0}\hat{U}(B_\mathbf{u})|\mathbf{0}> = \hat{U}(B_\mathbf{u})\left[\frac{\hat{\mathbf{P}}+\mathbf{u}\hat{P}^0}{\hat{P}^0+\mathbf{u}\cdot\hat{\mathbf{P}}}\right]|\mathbf{0}> = \hat{U}(B_\mathbf{u})\mathbf{u}|\mathbf{0}> = \mathbf{u}\hat{U}(B_\mathbf{u})|\mathbf{0}> . \tag{5.6}$$

Thus,

$$\hat{U}(B_\mathbf{u})|\mathbf{0}> = |\mathbf{u}> . \tag{5.7}$$

Consequently, the $|\mathbf{u}>$ can be found among the SP states for unstable quantons only on the *non*-instantaneous hyperplanes, and with different orientation of the hyperplane for every distinct velocity eigenvalue. Indeed, the 4-velocity eigenvalue, $v^\mu = (v^0, \mathbf{v}) = (1, \mathbf{u})/[1-\mathbf{u}^2]^{1/2}$, can be identified with the normal, $\eta^\mu$, to the hyperplane in question and (5.5) can be read as,

$$\hat{P}^\mu\,|\mathbf{u}> = v^\mu \sqrt{\hat{P}^2}\,|\mathbf{u}> = v^\mu (v\hat{P})\,|\mathbf{u}> , \tag{5.8}$$

synonymous, within the SP-momentum eigenstates, to

$$\hat{P}^\mu\,|\,p=0;\eta,\tau> = \eta^\mu (\eta\hat{P})\,|\,p=0;\eta,\tau> . \tag{5.9}$$

Thus all the velocity eigenstates among unstable quanton, SP states are SP-momentum eigenstates with the eigenvalue, $p = 0$. They are on hyperplanes orthogonal to their 4-velocity eigenvalues. Consequently, the velocity eigenstates do not span the SP-states for any hyperplane and a superposition of velocity eigenstates with distinct velocity eigenvalues would be a superposition of SP states on distinct, intersecting, hyperplanes and would, consequently, not be, itself, an SP state on any hyperplane! This completely undermines the adequacy of the velocity eigenstates as basis vectors for unstable quantons.

On the basis of the preceding we can assert the proportionality,

$$|\mathbf{u}> = |v> \propto |p = 0; \eta = v, \tau = 0>. \tag{5.10}$$

If we choose the invariant normalization,

$$<\mathbf{u}'|\mathbf{u}> = <v'|v> = v^0 \delta^3(\vec{v}' - \vec{v}) = (1 - \mathbf{u}^2)^2 \, \delta^3(\mathbf{u}' - \mathbf{u}), \tag{5.11}$$

then the proportionality can be shown to be replaceable by the equality (see **Appendix 2**),

$$|\mathbf{u}> = |v> = |p = 0; \eta = v, \tau = 0><(\hat{P}^2)^{-3/2}>^{-1/2}. \tag{5.12}$$

Among those who have worked with these velocity eigenstates, the survival amplitude, when considered, is defined by,

$$<\mathbf{u}'|\exp[-(i/\hbar)\hat{P}^0 ct]|\mathbf{u}> = (1 - \mathbf{u}^2)^2 \, \delta^3(\mathbf{u}' - \mathbf{u}) I_\mathbf{u}(t), \tag{5.13}$$

This definition is motivated by the (erroneous, from our SP perspective) interpretation of the velocity eigenstate as describing an unstable quanton in an ISP state, rather than an SP state on a non-instantaneous hyperplane (see also [Ste 08]). Nevertheless, it is a decay amplitude of sorts. Let's see what we get.

From the boost considered in (5.6) we have,

$$<\mathbf{u}'|\exp[-(i/\hbar)\hat{P}^0 ct]|\mathbf{u}> = <\mathbf{u}'|\hat{U}(B_\mathbf{u})\hat{U}(B_\mathbf{u})^\dagger \exp[-(i/\hbar)\hat{P}^0 ct]\hat{U}(B_\mathbf{u})|0>$$

$$= <B_\mathbf{u}^{-1}\mathbf{u}'|\exp\left[-(i/\hbar)\frac{\hat{P}^0 + \mathbf{u}\cdot\hat{\mathbf{P}}}{\sqrt{1-\mathbf{u}^2}}ct\right]|0> = <B_\mathbf{u}^{-1}\mathbf{u}'|\exp\left[-(i/\hbar)\frac{\hat{P}^0}{\sqrt{1-\mathbf{u}^2}}ct\right]|0>$$

$$= \delta^3(B_\mathbf{u}^{-1}\mathbf{u}')I_0(\frac{t}{\sqrt{1-\mathbf{u}^2}}) = (1-\mathbf{u}^2)^2 \delta^3(\mathbf{u}'-\mathbf{u})I_0(\frac{t}{\sqrt{1-\mathbf{u}^2}}). \tag{5.14}$$

The radical under the time variable in the survival amplitude guarantees the accelerated decay process and, consequently, that the lifetime will be contracted by just that radical factor, $\sqrt{1-\mathbf{u}^2}$. How are we to understand this? (**Fig.2**)

As we have seen, a 4-velocity eigenstate with eigenvalue, $v^\mu$, is an SP state on a hyperplane orthogonal to $v^\mu$. The SP-momentum in this state is $p = 0$ and, consequently, in a reference frame in which the SP hyperplane is instantaneous, the lifetime, which we recall depends only on $p^2$ (see (4.10)), is the lifetime at rest, $T_0$. If we then represent the average decay occurrence by a parallel hyperplane to the future of the SP hyperplane by the *proper time* $T_0$ in the direction $v^\mu$, we note that the *time* interval between the two parallel hyperplanes is just $T_0/v^0 = \sqrt{1-\mathbf{u}^2}\, T_0$, a *contraction* of $T_0$. There is an inclination to claim that the *time* interval we *should* consider is that between points on the parallel hyperplanes connected by the direction normal, $v^\mu$. *That* time interval is the standardly dilated, $v^0 T_0 = T_0 / \sqrt{1-\mathbf{u}^2}$. This claim, however, is incorrect. The quanton is not located at any point on either hyperplane, but, having $p = 0$, is over the hyperplane entire. Only the *time* interval *between the hyperplanes* is meaningful. This discussion should be compared with that in **4** on the lifetime dilation for $p \neq 0$.

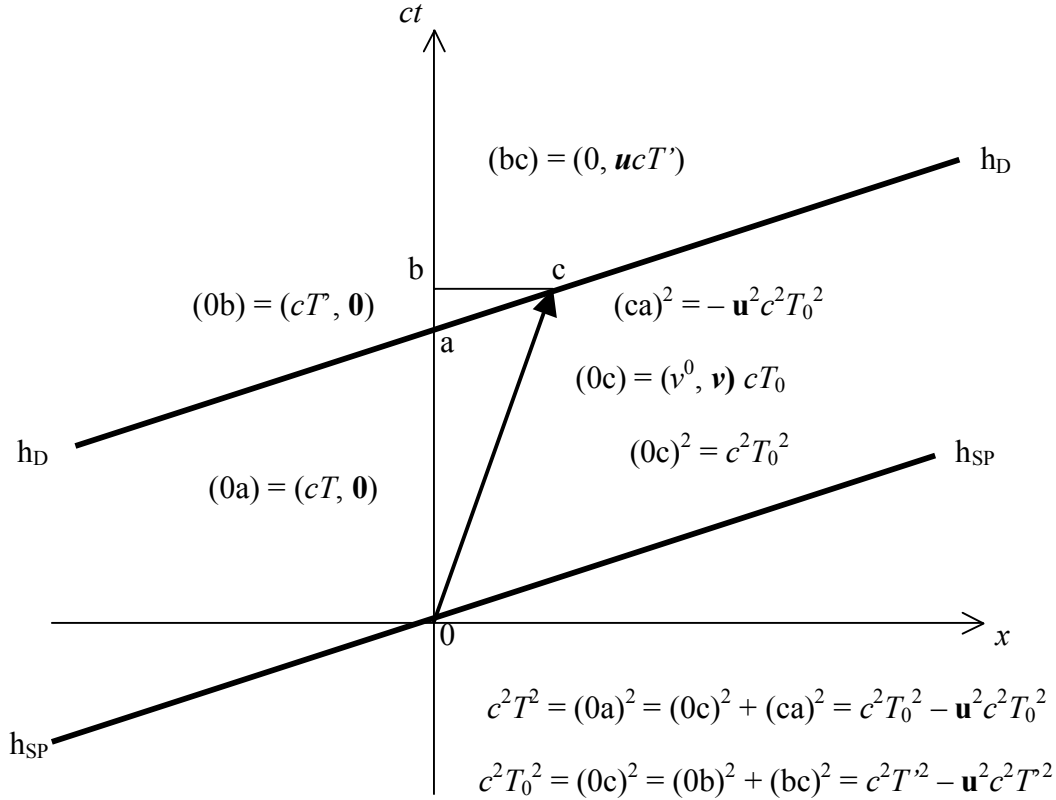

**Fig. 2:** Hyperplane $h_{IP}$ is the SP hyperplane of the velocity eigenstate with 4-velocity $v^\mu$. Hyperplane $h_D$ is the parallel hyperplane on which the decay occurs on average. The 4-vector, (0c), is orthogonal to these hyperplanes and has a magnitude given by the at-rest lifetime, $T_0$. The *time* interval, $T$, separating the hyperplanes is the magnitude of the 4-vector, (0a). $T'$ is the time interval separating the *non*-events, 0 and c. The 4-vectors, (bc) and (ca), are space-like with negative squared magnitudes. Applying the Lorentzian-Pythagorean theorem to the 'right triangles' 0bc and 0ca we find that $T_0$ is contracted to $T$ and dilated to $T'$.

One might still wonder whether an interpretation of the velocity eigenstates as instantaneous unstable quanton states might be mounted from the ground up if we do not insist in viewing such states from our hyperplane dependent, SP perspective. This has been done by A. Bohm and collaborators, [Bo 97, 00, 03], [Ta 08], but at the price of working in an extended state space allowing complex eigenvalues for the momenta, energies and masses of the corresponding operators and accepting fundamental temporal irreversibility. Aside from this option and the problem of interpreting the Shirokov lifetime contraction, it is easy to show, within a standard state space, the inconsistency of an ISP interpretation of a velocity eigenstate with non-zero eigenvalue.

Paraphrasing an argument made in **2**; an instantaneous, unstable quanton, non-zero velocity eigenstate at a particular time should remain such under arbitrary spatial displacement. By superimposing such spatial displacements of a single such velocity eigenstate we do not lose either the instantaneous or the velocity or the unstable quanton aspect of the state. But with *appropriate* superpositions we can build momentum eigenstates. These are then *joint* velocity-momentum eigenstates with, in general, non-zero eigenvalues. They are, therefore, also energy eigenstates and rest mass eigenstates. Thus they are *not* unstable quanton states; a contradiction! We avoid the contradiction by denying instantaneity for non-zero velocity eigenstates; Bohm et al avoid the contradiction by admitting energy eigenstates with complex eigenvalues.

**6. Field theory two-point VEVs:** We return to quantum fields to examine the contribution unstable quantons can make to the vacuum expectation values of products of two fields, traditionally called *two point functions*. Whereas in **3** the relations we obtained held for any field, $\hat{\phi}(x)$, here we will be concerned only with fields for which the contribution to the sum over intermediate states from pairs of stable quantons in the two-point functions is dominated by the principle decay products of our unstable quanton. This means that if, $\hat{\Pi}_2^{(\pm)}$, are the projection operators onto asymptotic incoming (-) or outgoing (+) scattering states for all pairs of stable quantons, then,

$$<\Omega|\hat{\phi}(x')\hat{\Pi}_2^{(\pm)}\hat{\phi}(x)|\Omega> \approx \int d^4q\, d^2\mathbf{n} <\Omega|\hat{\phi}(x')|q,\mathbf{n},(\pm)><q,\mathbf{n},(\pm)|\hat{\phi}(x)|\Omega>, \quad (6.1)$$

where the $|q, \mathbf{n}, (\pm)>$ describe asymptotic scattering states of the two stable decay products with total 4-momentum, $q$, and rest frame, 3-momentum direction unit vector, $\mathbf{n}$. These 2-quanton scattering states are related to the 4-momentum eigenstates, $|q>$, contributing to the SP-momentum eigenstates for the unstable quanton as indicated in (4.7) by a similar dominance relation,

$$|q> \approx \frac{1}{\sqrt{4\pi}} \int d^2\mathbf{n}\, |q,\mathbf{n},(\pm)>. \quad (6.2)$$

Expressing the action of the field on the vacuum as,

$$\hat{\Pi}_2^{(\pm)}\hat{\phi}(x)|\Omega> \approx \int d^4q\, d^2\mathbf{n}\, |q,\mathbf{n},(\pm)> f(q^2)\exp[(i/\hbar)qx], \quad (6.3)$$

we find,

$$<\Omega|\hat{\phi}(x')\hat{\Pi}_2^{(\pm)}\hat{\phi}(x)|\Omega> \approx 4\pi \int d^4q\, |f(q^2)|^2 \exp[(i/\hbar)q(x-x')]. \quad (6.4)$$

It will be useful in the following to express the right hand side of (6.4) in two different forms. We will use (2.18) in the form,

$$d^4q = d^3_\eta q\, d(\eta q) = \frac{d^3_\eta q}{2\eta q} 2\eta q\, d(\eta q) = \frac{d^3_\eta q}{2\eta q} d(q^2) = \frac{d^3_\eta p}{2\sqrt{\mu^2 - p^2}} d\mu^2, \qquad (6.5)$$

where, $p = q - \eta(\eta q)$, and $\mu^2 = q^2$. The right hand side of (6.4) is, on one hand, equal to,

$$4\pi \int d^3_\eta p \exp[\frac{i}{\hbar} p(x-x')] \left( \int \frac{d\mu^2}{2\sqrt{\mu^2 - p^2}} |f(\mu^2)|^2 \exp[\frac{i}{\hbar}\sqrt{\mu^2 - p^2}(\eta x - \eta x')] \right). \qquad (6.6)$$

By reordering the integrations, it is also equal to,

$$4\pi \int d\mu^2 |f(\mu^2)|^2 \Delta_\mu(x - x'), \qquad (6.7)$$

where,

$$\Delta_\mu(x - x') = \int \frac{d^3_\eta p}{2\sqrt{\mu^2 - p^2}} \exp[\frac{i}{\hbar}(p + \eta\sqrt{\mu^2 - p^2})(x - x')], \qquad (6.8)$$

is a standard causal Green's function, or propagator (the seeming $\eta$ dependence in (6.6) and (6.8) is illusory and an artifact of the mode of expression).

We are going to compare these results with the quantity,

$$<\Omega|\hat{\phi}(x')\hat{\Pi}(\eta,\eta x')\hat{\Pi}(\eta,\eta x)\hat{\phi}(x)|\Omega>, \qquad (6.9)$$

in which *two* projection operators onto the state spaces of the unstable quanton,

$$\hat{\Pi}(\eta,\tau) := \int d^3_\eta p\, |p;\eta,\tau><p;\eta,\tau|, \qquad (6.10)$$

have been inserted between the field operators. Why two and why the choice of $\tau$ values, $\eta x$ and $\eta x'$?

For stable quantons there is only one projection operator onto the state space for a single quanton. To insert it twice would be equivalent to inserting it once. For our unstable quanton, however, there is a distinct projection operator for each hyperplane on which the quantons are undecayed. We wish to examine the degree to which the unstable quanton states can exhaust the intermediate state contribution to the field's two point function. But for the unstable quanton the propagation between the field points $x$ and $x'$ is going to involve the possibility of decay. By inserting the projectors as indicated in (6.9) we focus on the intermediate state contribution from the unstable quanton created 'at x' on the ($\eta$, $\eta x$) hyperplane, which contains $x$, and propagating, with the possibility of decay, to annihilation 'at x'' on the ($\eta$, $\eta x'$) hyperplane, which contains $x'$. Using (3.8) and (6.10), we have,

$$\hat{\Pi}(\eta,\eta x')\hat{\Pi}(\eta,\eta x) = \int d_\eta^3 p \, | p;\eta,\eta x' > I_p(\eta x' - \eta x) < p;\eta,\eta x |. \qquad (6.11)$$

Inserting this in (6.9) and using (4.7) and (6.2, 3) to obtain,

$$< p;\eta,\eta x | \hat{\phi}(x) | \Omega > \approx \sqrt{4\pi} \int \frac{d\mu^2}{\sqrt{2\sqrt{\mu^2 - p^2}}} r^*(\mu^2) f(\mu^2) \exp[(i/\hbar)px], \qquad (6.12)$$

we find,

$$< \Omega | \hat{\phi}(x') \hat{\Pi}(\eta,\eta x') \hat{\Pi}(\eta,\eta x) \hat{\phi}(x) | \Omega >$$

$$\approx 4\pi \int d_\eta^3 p \exp[(i/\hbar)p(x - x')] I_p(\eta x' - \eta x) \left| \int \frac{d\mu^2}{\sqrt{2\sqrt{\mu^2 - p^2}}} r^*(\mu^2) f(\mu^2) \right|^2. \qquad (6.13)$$

If we now combine this with (6.4), with the right hand side expressed as in (6.6) and also employing (4.9, 11), we find,

$$< \Omega | \hat{\phi}(x') \{\hat{\Pi}_2^{(\pm)} - \hat{\Pi}(\eta,\eta x') \hat{\Pi}(\eta,\eta x)\} \hat{\phi}(x) | \Omega >$$

$$\approx 4\pi \int d_\eta^3 p \exp[\frac{i}{\hbar} p(x-x')] \left\{ \left( \int d\mu^2 \left| \frac{f(\mu^2)}{\sqrt{2\sqrt{\mu^2 - p^2}}} \right|^2 \exp[\frac{i}{\hbar}\sqrt{\mu^2 - p^2}(\eta x - \eta x')] \right) \right.$$

$$\left. - \left( \int d\mu'^2 \, |r(\mu'^2)|^2 \exp[\frac{i}{\hbar}\sqrt{\mu'^2 - p^2}(\eta x - \eta x')] \right) \left| \int d\mu^2 r^*(\mu^2) \frac{f(\mu^2)}{\sqrt{2\sqrt{\mu^2 - p^2}}} \right|^2 \right\}. \qquad (6.14)$$

The reason for writing this complex equation out in so much detail is to enable the reader to see by inspection that (6.14) would vanish if only we could choose,

$$f(\mu^2) = \sqrt{2\sqrt{\mu^2 - p^2}} \, r(\mu^2). \qquad (6.15)$$

Of course, this choice is not permitted since the function, $f$, introduced in (6.3), depends only on one variable, $\mu^2$, not on $\mu^2$ and $p^2$. The best we can do is to choose,

$$f(\mu^2) = \sqrt{2\mu}\, r(\mu^2), \tag{6.16}$$

which makes the contribution to the integral over $p$, in (6.14), vanish at $p = 0$, but not elsewhere. This looks to be the best one can do towards having the intermediate unstable quantons dominate the two point VEV for a *local* field.

This result provokes the speculation of the possible value of introducing, for unstable quantons, a *minimally non-local, hyperplane dependent* field, $\hat{\phi}(\eta,x)$, which satisfies, in contrast to (6.3),

$$\hat{\Pi}_2^{(\pm)}\hat{\phi}(\eta,x)|\Omega> \approx \int d^4q\, d^2\mathbf{n}\,|q,\mathbf{n},(\pm)> \sqrt{2\eta q}\, r(q^2)\exp[(i/\hbar)qx]$$

$$\approx \sqrt{4\pi}\int d_\eta^3 p\, |p;\eta,\eta x>\exp[(i/\hbar)px] = \hat{\Pi}(\eta,\eta x)\hat{\phi}(\eta,x)|\Omega>. \tag{6.17}$$

We will leave the matter there.

**7. The last case:** In **2** we examined inner products of SP momentum eigenstates, $<p';\eta',\tau'|p;\eta,\tau>$, with $\eta' = \eta$ and $\tau' = \tau$. In **3** we considered $\tau' \neq \tau$. In **5** we examined inner products of velocity eigenstates, $<v'|v>$, which is the case of SP inner products with $p' = p = 0$ and $\eta' = v' \neq v = \eta$. In this section we look briefly at the remaining case in which, $\eta' \neq \eta$ and $0 \neq p'$ or $p \neq 0$ or both.

In this case we have, using (4.7, 8),

$$<p';\eta',\tau'|p;\eta,\tau>$$

$$= 2\int d^4q\, \delta_\eta^3(p'-q)\delta_\eta^3(p-q)\sqrt{(\eta'q)(\eta q)}\,|r(q^2)|^2\exp[(i/\hbar)((\eta q)\tau - (\eta'q)\tau')]. \tag{7.1}$$

The delta functions appearing under the integral sign can be written as,

$$\delta_\eta^3(p-q) = \delta_{\eta',\eta}^2(p-q)\delta(e(p-q)), \tag{7.2a}$$

and,

$$\delta_\eta^3(p'-q) = \delta_{\eta',\eta}^2(p'-q)\delta(e'(p'-q)), \tag{7.2b}$$

where $\delta_{\eta',\eta}^2$, denotes a two-dimensional delta function in the space-like directions orthogonal to both $\eta$ and $\eta'$, while $e$ and $e'$ are dimensionless unit vectors lying in the $(\eta, \eta')$ plane and orthogonal, respectively, to $\eta$ and $\eta'$. More precisely,

$$e = \frac{\eta' - (\eta'\eta)\eta}{\sqrt{(\eta'\eta)^2 - 1}}, \tag{7.3a}$$

and

$$e' = \frac{\eta - (\eta'\eta)\eta'}{\sqrt{(\eta'\eta)^2 - 1}}. \tag{7.3b}$$

Remembering that $p$ and $p'$ are also orthogonal, respectively, to $\eta$ and $\eta'$, we can use (7.2a) in (7.1) to reduce the 4-dimensional integral, $\int d^4q \text{ - - - } = \int d_\eta^3 q \, d(\eta q) \text{ - - -}$ to the 1-dimensional integral, $\int d(\eta q) \text{ - - -}$, replacing $q$ everywhere by $p + \eta(\eta q)$ and replacing (7.2b) by

$$\delta_\eta^3(p' - q) = \delta_{\eta',\eta}^2(p' - p)\delta(e'(p' - p - \eta(\eta q))). \tag{7.4}$$

Substituting (7.3b) into the last delta function on the right side of (7.4), we find,

$$\delta(e'(p' - p - \eta(\eta q))) = \sqrt{(\eta'\eta)^2 - 1}\,\delta(\eta p' + (\eta'\eta)\eta' p + ((\eta'\eta)^2 - 1)\eta q)$$

$$= \frac{\delta\left(\eta q + \frac{(\eta p' + (\eta'\eta)\eta' p)}{(\eta'\eta)^2 - 1}\right)}{\sqrt{(\eta'\eta)^2 - 1}}. \tag{7.5}$$

The unexpected result here is that when $\eta \neq \eta'$ and $0 \neq p$ or $p' \neq 0$ or both, the delta functions in (7.1) permit, *at most, one* value of the 4-momentum, $q$, to contribute to the integral and thus, to the inner product. That value, assuming it lies within the allowed spectrum for $q$, determined by the spectral function, $r(q^2)$, must satisfy,

$$q = p + \eta(\eta q) = p_{\eta',\eta} - e(ep) + \eta(\eta q) = p' + \eta'(\eta' q) = p_{\eta',\eta} - e'(e' p') + \eta'(\eta' q), \tag{7.6}$$

where $p_{\eta',\eta}$ is, from (7.4), the shared component of $p$ and $p'$ orthogonal to the ($\eta$, $\eta'$) plane (**Fig. 3**). Solving for $\eta q$ and/or $\eta' q$ yields, as indicated by (7.5),

$$q = p - \eta \frac{\eta p' + (\eta'\eta)\eta' p}{(\eta'\eta)^2 - 1} = p' - \eta' \frac{\eta' p + (\eta'\eta)\eta p'}{(\eta'\eta)^2 - 1} = p_{\eta',\eta} - \frac{\eta'(\eta' p) + \eta(\eta p')}{(\eta'\eta)^2 - 1}, \tag{7.7}$$

where $p_{\eta',\eta}$ is the common part of $p$ and $p'$ that is orthogonal to both $\eta$ and $\eta'$, as insured by the two-dimensional delta function, $\delta_{\eta',\eta}^2$, in (7.4). The resulting value for the 4-momentum, $q$, will be physically accessible only if both, $\eta p'$ and $\eta' p$ are negative and sufficiently large. The final result for the inner product, (7.1), is

$$<p';\eta',\tau'\,|\,p;\eta,\tau> = 2\delta_{\eta',\eta}^2(p' - p)\frac{\sqrt{(\eta p' + (\eta'\eta)\eta' p)(\eta' p + (\eta'\eta)\eta p')}}{[(\eta'\eta)^2 - 1]^{3/2}}$$

$$\left| r\left( (p_{\eta',\eta})^2 + \frac{(\eta(\eta p') + \eta'(\eta' p))^2}{((\eta'\eta)^2 - 1)^2} \right) \right|^2 \exp\left[ \frac{i}{\hbar} \frac{(\eta p' + (\eta'\eta)\eta' p)\tau - (\eta' p + (\eta'\eta)\eta p')\tau'}{1 - (\eta'\eta)^2} \right], \qquad (7.8)$$

a result displaying no hint of decay in its dependence on $\tau$ and $\tau'$, but a pure oscillatory phase factor dependence instead!

In fact, this result makes perfect physical and geometrical sense. The unstable quantons are in SP momentum eigenstates with no preferential *location* on the respective *intersecting* hyperplanes. Varying $\tau$ and/or $\tau'$, while changing the *hyperplanes*, does not change the global relationship between them. It merely slides the intersection between them along each of them, providing no geometrical/physical basis for a change in the *magnitude* of the inner product. Very different from the case of parallel hyperplanes in

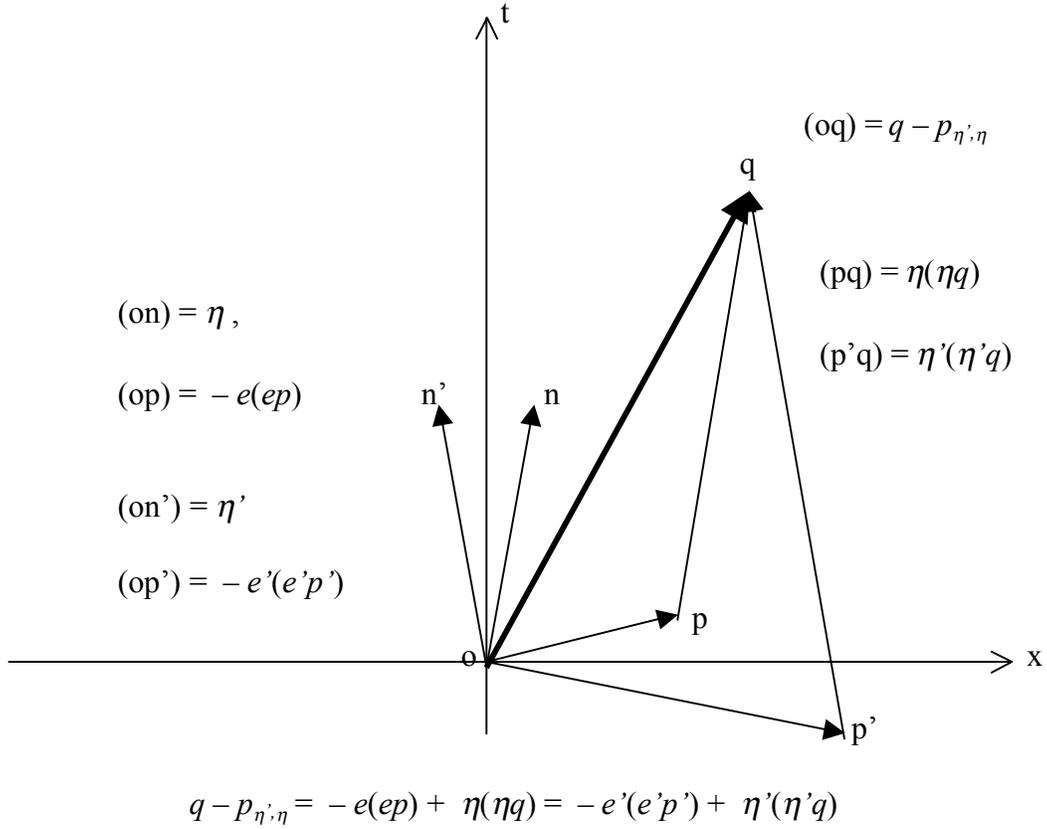

$$q - p_{\eta',\eta} = -e(ep) + \eta(\eta q) = -e'(e'p') + \eta'(\eta' q)$$

**Fig. 3:** Space-time diagram of the ($\eta$, $\eta'$) plane and the relationships among the 4-vectors, $\eta$, $\eta'$, $-e(ep)$, $-e'(e'p')$, $\eta(\eta q)$, $\eta'(\eta' q)$ and $q - p_{\eta',\eta}$. The perspective is that of a reference frame in which $\eta'$ and $\eta$ have equal and opposite space components.

which varying $\tau$ and/or $\tau'$ changes the time-like distance between the hyperplanes, thereby allowing for decay. To see decay-like behaviour on hyperplanes intersecting the no-decay hyperplane of a single parent, one must consider SP states that are space-like *localized* to some degree on either the no-decay hyperplane, the intersecting hyperplanes, or both. This remark is relevant to the treatment of Stefanovich [Ste 08].

As a parting shot we note that, notwithstanding the odd form of (7.8) and the different forms of (5.11) and (2.12) above (see also (A2.3) below), it is easy to show that,

$$\hat{\Pi}(\eta',\tau')\hat{\Pi}(\eta,\tau)$$

$$= \int d^4q \,|q-\eta'(\eta'q);\eta',\tau'> 2\sqrt{(\eta'q)(\eta q)} \,|r(q^2)|^2 \,\exp[\frac{i}{\hbar}q(\eta\tau-\eta'\tau')]<q-\eta(\eta q);\eta,\tau|. \quad (7.9)$$

**Acknowledgement:** This research was motivated and initiated during a short term visit to Perimeter Institute, Waterloo, Ontario, in June of 2008. I wish to thank the Institute for its hospitality and stimulating atmosphere and my hosts, Owen Maroney and Harvey Brown, for the invitation to visit.

**Appendix 1:** (Derivation of (4.13)) From (4.11, 12) we have, where $\varepsilon_p(\mu) := \sqrt{\mu^2 - p^2}$,

$$cT_p = \int_0^\infty d\tau \left| \int d\mu \, \sigma(\mu) \exp[-(i/\hbar)\varepsilon_p(\mu)\tau] \right|^2$$

$$= \iint d\mu' d\mu \, \sigma(\mu')\sigma(\mu) \left[ \int_0^\infty d\tau \exp[(i/\hbar)(\varepsilon_p(\mu') - \varepsilon_p(\mu))\tau] \right]$$

$$= \iint d\mu' d\mu \, \sigma(\mu')\sigma(\mu)(i\hbar) \left( \frac{P}{\varepsilon_p(\mu') - \varepsilon_p(\mu)} - i\pi \delta(\varepsilon_p(\mu') - \varepsilon_p(\mu)) \right)$$

$$= \pi\hbar \iint d\mu' d\mu \, \sigma(\mu')\sigma(\mu) \delta(\varepsilon_p(\mu') - \varepsilon_p(\mu)) = \pi\hbar \iint d\mu' d\mu \, \sigma(\mu')\sigma(\mu) \frac{\delta(\mu'-\mu)}{|\partial\varepsilon_p(\mu)/\partial\mu|}$$

$$= \pi\hbar \int d\mu \frac{\sigma(\mu)^2}{|\partial\varepsilon_p(\mu)/\partial\mu|} = \pi\hbar \int d\mu \, \sigma(\mu)^2 \frac{\sqrt{\mu^2 - p^2}}{\mu}, \quad (A1.1)$$

the desired result.

**Appendix 2:** (Derivation of (5.12)) Writing, $|0;v,0> := |p=0;\eta=v,\tau=0>$, we have, from (4.7) and (4.8),

$$<0;v',0|0;v,0> = 2\int d^4q\, \delta_v^3(q)\sqrt{v'q}\,|r(q^2)|^2\,\sqrt{vq}\,\delta_v^3(q). \tag{A2.1}$$

From (2.13) this becomes,

$$2\int dq^0\, \frac{q^0}{(v^0)^3}\delta^3\left(\left(\frac{\mathbf{v'}}{v'^0}-\frac{\mathbf{v}}{v^0}\right)q^0\right)\left|r\left(\left(\frac{q^0}{v^0}\right)^2\right)\right|^2 = (v^0)^{-4}\,\delta^3(\mathbf{u'}-\mathbf{u})\int \frac{dq^0}{(q^0)^2}v^0\,2\left|r\left(\left(\frac{q^0}{v^0}\right)^2\right)\right|^2$$

$$=(1-\mathbf{u}^2)^2\,\delta^3(\mathbf{u'}-\mathbf{u})\int d\mu\, \sigma(\mu)\mu^{-3} = v^0\delta^3(\mathbf{v'}-\mathbf{v}) < (\hat{P}^2)^{-3/2}>, \tag{A2.2}$$

or

$$<0;v',0|0;v,0> = v^0\delta^3(\mathbf{v'}-\mathbf{v})<(\hat{P}^2)^{-3/2}> = <v'|v><(\hat{P}^2)^{-3/2}>. \tag{A2.3}$$

From this and (5.10), (5.12) follows.